\documentclass[11pt]{JHEP3}

\input amssym.def
\input amssym.tex

\newcommand{\dalpha}{{\dot{\alpha}}}
\newcommand{\dbeta}{{\dot{\beta}}}

\newcommand{\N}{{\cal N}}

\newcommand{\half}{{{\textstyle\frac{1}{2}}}}

\newcommand{\be}{\begin{equation} }
\newcommand{\ee}{\end{equation} }
\newcommand{\ba}{\begin{array}}
\newcommand{\ea}{\end{array}}

\newcommand{\z}{\hat{z}}
\newcommand{\w}{\hat{w}}

\newcommand{\x}{\hat{x}}
\newcommand{\y}{\hat{y}}
\newcommand{\htheta}{\hat{\theta}}
\newcommand{\btheta}{\bar{\theta}}
\newcommand{\hbtheta}{\hat{\bar{\theta}}}
\newcommand{\hvartheta}{\hat{\vartheta}}
\newcommand{\heta}{\hat{\eta}}
\newcommand{\hpartial}{\hat{\partial}}

\def\lrOmega{{\stackrel{\leftrightarrow}{\Omega}}}
\def\rpartial{{\stackrel{\leftarrow}{\partial}}}
\def\lpartial{{\stackrel{\rightarrow}{\partial}}}

\def\O{{\cal O}}

\def\Q{{\cal Q}}

\def\tr{{\rm tr}}
\def\Tr{{\rm Tr}}
\def\I_M{{I_{\scriptscriptstyle M\times M}}}

\def\N{{\cal  N}}

\title{Superfield theory and supermatrix model}

\author{Jeong-Hyuck Park\\
Physics Department, Sungkyunkwan University\\
Chunchun-dong, Jangan-gu, Suwon 440-746, Korea\\
Electronic correspondence : \email{jhp@newton.skku.ac.kr}}

\abstract{We study the noncommutative superspace of arbitrary dimensions in a systematic way. Superfield theories on a  noncommutative superspace can be formulated in two folds, through the star product formalism and in terms of  the supermatrices.  We elaborate the duality  between them by constructing  the isomorphism explicitly and relating the superspace integrations of the star product Lagrangian or the superpotential to the traces of the supermatrices.  We show there exists  an interesting fine tuned commutative limit where the duality can be  still maintained.  Namely on the commutative superspace too, there  exists a   supermatrix model description for the superfield theory. We interpret the result in the context of the wave particle duality. The dual particles for the superfields in even and odd spacetime dimensions are D-instantons and D0-branes respectively to be   consistent with the T-duality.}


\preprint{hep-th/0307060}

\begin{document}
\section{Introduction and summary}
The wave particle duality was  the key ingredient for the birth of the quantum theory. In the standard quantum field theory, the field  represents the corresponding wave while the notion of the particle arises after the second quantization of the classical field.  Remarkably in string/M-theory the wave particle duality is often  manifest   even at the classical level. The BFSS matrix model   is essentially  a description of the M-theory in terms of the infinitely many  D-particles \cite{BFSS}, while  the IKKT matrix model formulates IIB superstring theory through the dynamics of the D-instantons \cite{Ishibashi:1996xs}. Expanding the \textit{matrix} models around the D-brane solution one can derive the  supersymmetric Yang-Mills \textit{field} theory action or the worldvolume description of the D-branes (see e.g. \cite{Taylor:2001vb,Douglas:2001ba,JHPpp}). One  characteristic feature of the matrix models is that they incorporate  the identical nature of the particles since the permutation symmetry  for the labelling of   the  particles is taken as the gauge symmetry \cite{Poly,JHP}. \\

The close relation between the field theory and the matrix model was first studied  in the early eighties by Eguchi and Kawai \textit{et al.} \cite{largeN1,largeN2}, and more recently by  Dijkgraaf and Vafa  showing that the effective superpotential  in the  four dimensional $\N=1$ supersymmetric gauge theory can be read off from the associated  matrix model computation \cite{DV}. \\

The purpose of the present paper is to show directly that there exits a dual  supermatrix model for the low energy description  of any superfield theory.  Our scheme is first to introduce the noncommutativity on the superspace  \cite{Klemm:2001yu,Ooguri:2003qp,Seiberg:2003yz,deBoer:2003dn,Kawai:2003yf,Chepelev:2003ga,Britto:2003aj,
Terashima:2003ri,Hatsuda:2003ry,Ferrara} 
replacing all the ordinary products of the superfields by the generalized Moyal-Weyl star products. The theory is then equivalent to the operator formalism on a noncommutative superspace or
 the supermatrix model description.  A remarkable fact is that, contrary to the nonsupersymmetric case, the equivalence between the superfield theory and the supermatrix model  is still maintained  in a certain  commutative limit. Thus, turning off the noncommutativity in a fine tuned manner we are able to recover the original commutative superfield theory  accompanied by   the  dual supermatrix model.\footnote{After submitting the first version of our paper, we learned that \cite{Kawai:2003yf} contains the similar idea  but  differs in details.}\\

The main content of our paper is to establish  precisely the  equivalence  between the star product formalism and the supermatrix description.  The organization of the paper is as follows.

In section \ref{superspace},  we first  consider the most general noncommutative relations on the superspace. In contrast  to the work by Seiberg on the  noncommutative  Euclidean superspace \cite{Seiberg:2003yz}, we have in mind the study of the real action in the Minkowskian spacetime so that we consider all the possible noncommutativity among all the coordinates on the superspace.  In particular, we first consider the real basis for the superspace coordinates. Namely, for the complex Grassmann coordinates in the four dimensional chiral superspace, we decompose them into the real ones. Analyzing  the real basis we are able to obtain the general unitarity representations of the  noncommutative superspace.

In section \ref{star}, we show in detail how the star product formalism on a commutative superspace realizes  the noncommutative superspace.  Explicitly through the generalized Weyl ordering prescription, we construct  the isomorphism between the star product on the commutative superspace and the operator product on  the noncommutative superspace.

In section \ref{smatrix}, we study the supermatrix representation of the operators on the noncommutative superspace. To do so we first find the `canonical' form for the  noncommutativity parameters under the linear transformations of the super-coordinates. We show that the  generic  noncommutative superspace decomposes into two sub-superspaces which commute each other. One consists of   the bosonic and fermionic  harmonic oscillators  having the nondegenerate noncommutativity parameters, while on the other sub-superspace, the only possible noncommutative relation is given by the commutator between the even and odd coordinates.  Surely  the bosonic harmonic oscillators are represented by matrices of the infinite size and the fermionic oscillators or the Clifford algebra have the gamma matrix representation.  For the second sub-superspace we restrict on the commutative sub-superspace and continue the analysis for simplicity.  The superfields on the noncommutative superspace are then represented by supermatrices. Depending on the statistics of the superfields the corresponding supermatrices have the different structures. In particular, the supermatrix representation of the ordinary Grassmann number is given by the ``gamma five" matrix. We also study the noncommutative chiral superspace by complexifying the odd coordinates. We point out that the anti-commutator between the odd coordinate and its complex conjugate must be nonzero in  the generic nontrivial cases. 

In section \ref{actioncomm}, we rewrite  the integration of the star product Lagrangians  over the superspaces as the traces of the supermatrix Lagrangians. The integration over the full superspace is essentially given by the supertrace, while the integration over the chiral  superspace involves taking trace of the superpotential matrix times a certain gamma matrix. We also present formulae which relate  the partition function and the correlation functions in the star product superfield theory to those in the  corresponding supermatrix model.   Finally we argue   that there exits  an interesting commutative limit where the duality between the superfield theory and the supermatrix model  does not break down. In particular, in the commutative limit, the kinetic part of the supermatrix model tends to diverge so that it is natural to use the saddle point analysis imposing the constraint on the supermatrices which sets the divergent part to vanish. Hence, the saddle point approximation  essentially gives the low energy  description.

In section \ref{examples}, we explicitly  study two examples, $\N=1,D=3$ nonlinear sigma model and $\N=1,D=4$ chiral model.  In general, turning on the noncommutativity on the superspace breaks the supersymmetry. However, we show that  for a certain `minimally' noncommutative superspace, the supersymmetry is not broken in the chiral model.

Section \ref{comments} contains the comment on  the nonexistence of the $\N=4,D=4$ superfield formalism, and the Appendix carries our  proof of the isomorphism between the star product on the commutative superspace and the operator product on the noncommutative superspace. 

\section{Noncommutative superspace\label{superspace}}
Before we start, it is useful to define the following commutator or anti-commutator for the superspace coordinates, $z^{A}$,
\begin{equation}
[z^{A},z^{B}\}_{\pm}\equiv z^{A}z^{B}\pm(-1)^{\#_{\!A}\#_{\!B}}z^{B}z^{A}\,, 
\end{equation}
where $\#_{\!A}$ is the $Z_{2}$-grading of the superspace so that   $\#_{\!A}=0$  for even or bosonic  $A$, and $\#_{\!A}=1$ for odd or fermionic  $A$.  The ordinary commutative superspace  satisfies simply, 
\begin{equation}
[z^{A},z^{B}\}_{-}\!=0\,. 
\label{com}
\end{equation}
Any deformation of the above relation defines a noncommutative superspace on which   the coordinates,  $\z^{A}$,  are  operators and satisfy  the noncommutative relation,
\begin{equation}
[\hat{z}^{A},\hat{z}^{B}\}_{-}\!=\Omega^{AB}\,. \label{noncom}
\end{equation}
The noncommutative parameter, $\Omega^{AB}$, is taken here to be constant. Since the left and right hand sides must agree  for the $Z_{2}$-grading, ${\#_{\!A}+\#_{\!B}}\sim{\#_{\Omega^{AB}}}~\mbox{:\,mod~}2$. Note also
\begin{equation}
\Omega^{AB}+(-1)^{\#_{A}\#_{B}}\Omega^{BA}=0\,.\label{OmegaOmega}
\end{equation}

Explicitly with the even and odd coordinates, $\hat{z}^{\mu}$, $\hat{z}^{\alpha}$,
\begin{equation}
\ba{lll} [\z^{\mu},\z^{\nu}]=i\Theta^{\mu\nu}\,,~~~&~~~[\z^{\mu},\z^{\alpha}]=i\Psi^{\mu\alpha}\,,
~~~&~~~\{\z^{\alpha},\z^{\beta}\}=S^{\alpha\beta}\,, \ea
\end{equation}
so that
\begin{equation}
\Omega=\left(\ba{cr}i\Theta~&~i\Psi\\
{}&{}\\
-i\Psi^{T}~&~S\ea\right)\,.\label{Omega}
\end{equation}
The superdeterminant of $\Omega$ is then\footnote{The last equality in Eq.(\ref{sdet}) comes from the identity, 
$\det(1+i\Theta^{-1}\Psi S^{-1}\Psi^{T})=\det^{-1}(1+iS^{-1}\Psi^{T}\Theta^{-1}\Psi)$, which can be easily shown  using $\ln(\det M)=\tr(\ln M)$.} 
\begin{equation}
\mbox{sdet}(\Omega)=\displaystyle{\frac{\det({i\Theta-\Psi S^{-1}\Psi^{T}})}{\det(
S)}=\frac{\det(i\Theta)}{\det({S+i\Psi^{T}\Theta^{-1}\Psi})}}\,. \label{sdet}
\end{equation}
Henceforth, otherwise mentioned explicitly, we consider the real basis for the superspace with the reality condition, $(\z^{A})^{\dagger}=\z^{A}$. Taking the real basis will enable us to analyze the most general unitary representations of the noncommutative superspace as done in section \ref{smatrix}. In particular, the four dimensional  chiral superspace will be treated as the complexification of the corresponding real superspace. In the real basis   the noncommutative parameters, $\Theta^{\mu\nu}$, $\Psi^{\mu\alpha}$, $S^{\alpha\beta}$ are all real.


\section{Star product formalism\label{star}}
In this section, we show in detail how the star product formalism on a commutative superspace realizes  the noncommutative superspace.  We first define a one to one map, $\O$, from the commutative superspace to the noncommutative superspace by  the generalized Weyl ordering,
\begin{equation}
\O(z^{A_{1}}z^{A_{2}}\cdots z^{A_{n}})\!\equiv\!
\z^{[A_{1}}\cdots\z^{A_{n}\}_{+}}\!=\!\displaystyle{\sum_{P=1}^{n!}\frac{(-1)^{N_{P}}\!\!\!}{n!}}~
\z^{{P_{1}}}\cdots\z^{{P_{n}}}\,, \label{Omap}
\end{equation}
where $P$ denotes the permutation of $(A_{1},\cdots ,A_{n})$ and $N_{P}$ counts the number of  permuted times only for the
odd  coordinates  in $(z^{A_{1}}\cdots z^{A_{n}})$. The star $(\!\star)$ product of the superfields on the commutative superspace reads with the noncommutative parameter, $\Omega^{AB}$,
\begin{equation}
f(z)\star g(z)\equiv f(z)\,\displaystyle{e^{\frac{1}{2}\lrOmega}}\,g(z)\,,
\end{equation}
where explicitly in the expansion of the exponential,
\begin{equation}
\left(\lrOmega\right)^{n}=\rpartial_{\!A_{1}}\rpartial_{\!A_{2}}\cdots\rpartial_{\!A_{n}}
(\Omega^{A_{n}B_{n}}\lpartial_{\!B_{n}})\cdots
(\Omega^{A_{2}B_{2}}\lpartial_{\!B_{2}})(\Omega^{A_{1}B_{1}}\lpartial_{\!B_{1}})\,,
\end{equation}
which is consistent with the fact that $\lrOmega$ is even. The right and left derivatives are related by
({\textit{cf.}}~\cite{Seiberg:2003yz})
\begin{equation}
f\rpartial_{A}=(-1)^{\#_{\!A}(\#_{\!f}+1)}\lpartial_{A} f\,,
\end{equation}
so that $z^{B}\rpartial_{A}=\lpartial_{A}z^{B}=\delta_{A}{}^{B}$.

The duality between the operator and the star product formalisms comes from the following crucial isomorphism,
\begin{equation}
\O(f)\O(g)=\O(f\star g)\,,\label{iso}
\end{equation}
which also manifests the associativity of the star product. The Appendix carries our proof on the isomorphism. 

The derivatives act on the noncommutative superspace as
\begin{equation}
\ba{ll}
\O(\partial_{A}f)=[\hat{\partial}_{A},\O(f)\}_{-}\,,~~&~~\hpartial_{A}\equiv(\Omega^{-1})_{AB}\z^{B}\,.\ea
\label{deriv}
\end{equation}
It is also worth to note
\begin{equation}
z^{A}f(z)=\half[z^{A},f(z)\}_{\star+}=\half\left(z^{A}\star f(z)+(-1)^{\#_{\!A}\#_{\!f}}f(z)\star z^{A}\right)\,.\label{orditostar}
\end{equation}

\section{Supermatrix  representation\label{smatrix}}
Under the linear transformation, $\z^{A}\rightarrow \z^{\prime A}=P^{A}{}_{B}\z^{B}$, the noncommutative parameter,
$\Omega$,
transforms\footnote{The transpose of the supermatrix is defined as 
$(P^{T})_{A}{}^{B}=(-1)^{\#_{A}(\#_{A}+\#_{B})}P^{B}{}_{A}$ \cite{Buchbinder:uq}. } to $\Omega^{\prime}=P\Omega P^{T}$.   ~Acting a proper transformation  one can obtain its `canonical' form.  Firstly,  for $\det(\Theta)\det(S)\neq 0$,  we
consider the following transformation which preserves the reality, $(\z^{\prime A})^{\dagger}=
\z^{\prime A}$,
\begin{equation}
P\!=\!\!\left(\ba{cc}~1~&0\\
{}&{}\\ ~0~&1+\displaystyle{\sum_{n=0}^{\infty}}\,a_{n}(i\Psi^{T}\Theta^{-1}\Psi S^{-1})^{n}
\ea\right)\!\!\left(\ba{cc}1~~~&~0\\
{}&{}\\\Psi^{T}\Theta^{-1}~~~&~1\ea\right),\label{defP}
\end{equation}
where the coefficients are  given by  $a_{0}=0$, $a_{1}=-\frac{1}{2}$ and
\begin{equation}
\ba{ll}
a_{n+1}=-a_{n}-\textstyle{\frac{1}{2}}\sum_{j=1}^{n}a_{j}(a_{n+1-j}+a_{n-j})\,,~~&~~\mbox{for~\,} n\geq 1\,.\ea
\end{equation}
From the Grassmann property of $\Psi$, the sum in Eq.(\ref{defP}) is actually a finite one. The resulting noncommutative parameter is then block diagonal, $\Omega^{\prime}={\small{\left(\ba{cc}i\Theta&0\\0& S\ea\right)}}$,  as the odd parameter, $\Psi$, is now  completely removed.  In the case of $\det(\Theta)\det(S)=0$ too,  one can apply the previous analysis to the  nondegenerate sector and  achieve the canonical form.   The final result for the generic cases  can be summarized as follows. Any noncommutative superspace,  $\{\z^{A}\}$, decomposes into  two sub-superspaces which  commute each other,
\begin{equation}
\ba{c} \{\z^{A}\}=\{\z^{M}=(\x^{\mu},\hvartheta^{a})\}\oplus\{\w^{\dot{K}}=(\y^{\dot{\mu}},\heta^{\dot{a}})\}\,,\\
{}\\
(\z^{M})^{\dagger}= \z^{M}\,,~~~~(\w^{\dot{K}})^{\dagger}= \w^{\dot{K}}\,,~~~~[\z^{M},\w^{\dot{K}}\}=0\,.\ea
\end{equation}
Each sub-superspace satisfies
\begin{equation}
\ba{lll} [\x^{\mu},\x^{\nu}]=i\theta^{\mu\nu}\,, ~~~&~~
\{\hvartheta^{a},\hvartheta^{b}\}=s^{ab}\,, ~~~&~~ [\x^{\mu},\hvartheta^{a}]=0\,,\\
{}&{}&{}\\
{}[\y^{\dot{\mu}},\y^{\dot{\nu}}]=0\,, ~~~&~~ \{\heta^{\dot{a}},\heta^{\dot{b}}\}=0\,, ~~~&~~
[\y^{\dot{\mu}},\heta^{\dot{a}}]=i\chi^{\dot{\mu}\dot{a}}\,,\ea
\end{equation}
where all the parameters are real, and $\theta^{\mu\nu}$, $s^{ab}$ are nondegenerate,  $\det(\theta)\det(s)\neq 0$. Furthermore, as $\theta^{\mu\nu}$ and $s^{ab}$ are  respectively anti-symmetric real and positive definite symmetric real matrices,  there exit real linear transformations which  take them to the canonical forms,
\begin{equation}
\ba{ll} \theta^{\mu\nu}=\mbox{diag}\left(\epsilon,\epsilon,\cdots,\epsilon\right)\,,~\epsilon=
{\left(\ba{rr}0&1\\-1&0\ea\right)}\,,~~&~~~s^{ab}=2\delta^{ab}\,.\ea
\end{equation}
Namely, $\x^{\mu}$'s satisfy the Heisenberg algebra, while $\hvartheta^{a}$'s satisfy the Euclidean Clifford
algebra. Their unitary irreducible representations are unique and very well known, the harmonic oscillator type infinite matrices and the gamma matrices respectively. As for the second sub-superspace, $\{\w^{\dot{K}}\}$, if its only noncommutative parameter vanishes, $\chi^{\dot{\mu}\dot{a}}=0$, the sub-superspace becomes simply a commutative one. However, for the  generic case, $\chi^{\dot{\mu}\dot{a}}\neq 0$, we do not know the unitary representation, and henceforth we restrict on the commutative  case, $\chi^{\dot{\mu}\dot{a}}=0$. Furthermore, without loss of generality for any spacetime dimensions, we take  the dimension of the noncommutative odd coordinates to be  even, $\hvartheta^{a}$, $1\leq a\leq N$, $N=\mbox{even}$.  For example,   for the $\N$-extended supersymmetry in  the Minkowskian four dimensions, $N=4\N$. As the gamma matrices are then taken from the Euclidean even dimensions, they can be set    Hermitian and off-block diagonalized. The irreducible supermatrix representation of the noncommutative superspace coordinates  as well as the bosonic and fermionic component fields, $\phi(\x)$, $\psi(\x)$, are given by the direct product of the harmonic oscillator matrix and  the gamma matrix,
\begin{equation}
\ba{ll}
\hvartheta^{a}\Longrightarrow 1\otimes\left(\ba{cc}0~&\rho^{a}\\
(\rho^{a})^{\dagger}&0\ea\right)\,,~~~~&~~~~~
\rho^{a}(\rho^{b})^{\dagger}+\rho^{b}(\rho^{a})^{\dagger}=s^{ab}\,,\\
{}&{}\\
\x^{\mu}\Longrightarrow \x^{\mu}\otimes 1\,,~~~~&~~~~~w^{\dot{K}}\Longrightarrow w^{\dot{K}}\otimes \left(\ba{lc}1&0\\
0&~~~(-1)^{\#_{\!\dot{K}}}\ea\right)\,,\\
{}&{}\\
\O(\phi(x))\Longrightarrow\phi(\x^{\mu})\otimes 1\,,~~~~&~~~~~
\O(\psi(x))\Longrightarrow\psi(\x^{\mu})\otimes\left(\ba{lr}1&0\\
0&-1\ea\right)\,.
\ea\label{zwmap}
\end{equation}
Thus, in general, the supermatrix representation of the superfield, $\hat{F}(\z)$, on the noncommutative superspace is    of the form,
\begin{equation}
\hat{F}(\z)
=\sum
~\hat{F}_{[\vec{m},g,r]\,[\vec{n},h,s]}\,|\vec{m},g,r\rangle\langle \vec{n},h,s|\,,
\end{equation}
where  $[\vec{m},g,r]$ forms {\textit{an}}  index of the supermatrix such that $\vec{m}=(m_{1},m_{2},\cdot\cdot)$ are the non-negative integers  for the harmonic oscillators, $g$ denotes the gamma matrix index, and  $r$ is from the preexisting gauge group, if any \cite{commentJHP}. Surely the symmetric products of the infinite matrices, $\x^{\mu}$, the anti-symmetric products of the gamma matrices and the gauge group generators give the basis for the supermatrix representation.   As the superfield is  an eigenstate of the $Z_{2}$-grading,   the block diagonal and off-block diagonal components in the corresponding supermatrix have the opposite  $Z_{2}$-gradings, and hence the name `supermatrix',
\begin{equation}
\ba{lll}
\mbox{bosonic~superfield}~&\Longleftrightarrow
&~\left(\ba{cc}\mbox{even}&\mbox{odd}\\
{}&{}\\
\mbox{odd}&\mbox{even}\ea\right):~\mbox{bosonic~supermatrix}\,,\\
{}&{}&{}\\
\mbox{fermionic~superfield}~&\Longleftrightarrow
&~
\left(\ba{cc}\mbox{odd}&\mbox{even}\\
{}&{}\\
\mbox{even}&\mbox{odd}\ea\right):~\mbox{fermionic~supermatrix}\,.\ea
\end{equation}
The coefficients of the supermatrix may depend on the commutative coordinates, $w^{\dot{K}}$, if any.  It is also worth to note  that for an arbitrary  Grassmann number, $\varepsilon$, 
\begin{equation}
\left[\left(\ba{cc}\varepsilon&~0\\
0&-\varepsilon\ea\right),\mbox{~supermatrix~}\right\}_{-}=0\,,\label{consistency}
\end{equation}
which is  consistent with $[\varepsilon,\mbox{\,superfield}\,\}_{-}=0$. \\

Complexifying the fermionic  coordinates,  the noncommutative chiral superspace reads,
$(\x^{\mu},\htheta^{\alpha},\hat{\bar{\theta}}{}^{\dalpha}=(\htheta^{\alpha})^{\dagger})$,
$1\leq\alpha\leq N/2$,  satisfying
\begin{equation}
\ba{llll}
[\x^{\mu},\x^{\nu}]=i\theta^{\mu\nu}\,,~&~~\{\htheta^{\alpha},\htheta^{\beta}\}
=c^{\alpha\beta}\,,~&~~
\{\htheta^{\alpha},\hbtheta{}^{\dbeta}\}=\tilde{c}^{\alpha\dbeta}\,,
~&~~ \{\hbtheta{}^{\dalpha},\hbtheta{}^{\dbeta}\}=\bar{c}^{\dalpha\dbeta}\,.\ea  \label{chiral}
\end{equation}
As shown above, in the supermatrix formulation of the noncommutative superfield theories, the fermionic coordinates are represented by the Euclidean $N$-dimensional gamma matrices so that for the complex odd coordinates, 
\begin{equation}
\ba{ccc}
\htheta^{\alpha}=P^{\alpha}{}_{a}\Gamma^{a}\,,~~&~~~\{\Gamma^{a},\Gamma^{b}\}=2\delta^{ab}\,,
~~&~~~
(\Gamma^{a})^{\dagger}=\Gamma^{a}\,,
\ea
\end{equation}
where $1\leq \alpha\leq N/2$, $1\leq a\leq N$ and  $P^{\alpha}{}_{a}$ is a complex number. It follows that 
$\tilde{c}^{\alpha\dbeta}=2P^{\alpha}{}_{a}(P^{\beta}{}_{a})^{\ast}$, etc.   In particular, this implies that one can not simply switch off $\tilde{c}^{\alpha\dbeta}$  as it would   imply $P^{\alpha}{}_{a}=\htheta^{\alpha}=0$. \\

It is worth to note that  the following  choice of $P^{\alpha}{}_{a}$ gives the conventional fermionic harmonic oscillators,
\begin{equation}
\begin{array}{lll}
P^{\alpha}{}_{a}=
\textstyle{\frac{1}{\sqrt{2}}}(\delta^{2\alpha-1}{}_{a}+i\delta^{2\alpha}{}_{a})\,,
&~\Longrightarrow~&
\tilde{c}^{\alpha\dbeta}=2\delta^{\alpha\dbeta}\,,~~~~c^{\alpha\beta}=0\,,~~~~
\bar{c}^{\dalpha\dbeta}=0\,,
\end{array}
\label{minimal}
\end{equation}
which we will call the `minimally noncommutative chiral superspace' henceforth. 
Unless  mentioned  otherwise, the  minimal form is not to be  assumed below.\\

In the Minkowskian four dimensions, with the standard convention, $x^{\mu}_{\pm}=x^{\mu}\pm i\theta\sigma^{\mu}\bar{\theta}$, the chiral superspace normally means $\{x_{+},\theta\}$ rather than $\{x,\theta\}$.  However, in that case the star product of chiral superfields does not preserve the chirality. Thus,  what we mean by the  chiral superspace is strictly  $\{x,\theta\}$ in the present paper which was also the case in  \cite{Kawai:2003yf}. We will come back to the chiral superspace   in  Section \ref{examples}.
\newpage
\section{Supermatrix action  and  the commutative limit\label{actioncomm}}
Now, with the map from the commutative superspace to the noncommutative superspace, (\ref{Omap}), (\ref{zwmap}),
the superfield action reads
\begin{equation}
\displaystyle{\int dx^{D}d\vartheta^{N}{\cal L}_{\star}}
=\displaystyle{\sqrt{(-2\pi i)^{D}\mbox{sdet}{\,\Omega}}\,\mbox{Tr}
\left[\Gamma_{\!\!\scriptscriptstyle{(\!N+1\!)}}\O({\cal
L}_{\star})\right]}\,,\label{action}
\end{equation}
where ${\cal L_{\star}}$ denotes the superfield Lagrangian equipped with the star product. 
$\Gamma_{\!\scriptscriptstyle{(\!N+1\!)}}$ is a
normalized matrix or the generalized ``gamma five",
\begin{equation}
\Gamma_{\!\scriptscriptstyle{(\!N+1\!)}}=\displaystyle{\sqrt{\frac{2^{N}}{\det{s}\,}}}\,
\hvartheta^{[N}\cdots\hvartheta^{2}\hvartheta^{1]}=
\sqrt{(-1)^{\frac{1}{2}N(N-1)}}\left(\begin{array}{rr}1&0\\0&-1\end{array}\right)\,,
\end{equation}
and hence,  $\mbox{Tr}\left[\Gamma_{\!\!\scriptscriptstyle{(\!N+1\!)}}~\cdot\,\right]$ is essentially $\sqrt{(-1)^{\frac{1}{2}N(N-1)}}$ times the supertrace.

The partition function in the superfield theory is  related to that of the supermatrix model,  
\begin{equation}
\displaystyle{\int \!{DF}\,e^{i\int\! dz^{D+N}{\cal L}_{\star}(F)}}
=\mbox{sdet}\!\!\left(\textstyle{\frac{\partial F}{\partial \hat{F}}}\right)\displaystyle{\int\! D\hat{F}\,
e^{i\sqrt{(-2\pi i)^{D}\mbox{sdet}{\,\Omega}}\,\mbox{Tr}
\left[\Gamma_{\!\scriptscriptstyle{(\!N+1\!)}}{\cal L}(\hat{F})\right]}}\,, 
\label{partition}
\end{equation}
where $\hat{F}=\O(F(z))$ is the supermatrix. Since $\O$ is a linear map 
the Jacobian is constant, and  we are able to take it out from the integration.  Furthermore, from the expression of  the partition function above, the correlators of  the superfields can be obtained   from the supermatrix correlators \cite{Langmann:2003cg}, 
\begin{equation}
\langle F(z_{1})F_(z_{2})\rangle=\langle \hat{F}_{LM}\hat{F}_{JK}\rangle\Lambda^{LM}(z_{1})\Lambda^{JK}(z_{2})\,,
\end{equation}
where $L,M$ denote the indices of the supermatrix, $\hat{F}$, while $\Lambda^{LM}(z)$ forms a basis for  the functions on the superspace satisfying $\O(\Lambda^{LM})_{JK}=\delta^{L}_{\,J}\,\delta^{M}_{~K}$ so that $F(z)=\hat{F}_{LM}\Lambda^{LM}(z)$ and $\O(\hat{F}_{LM}\Lambda^{LM})=\hat{F}$. \\

The conventional supersymmetry operator, $Q_{a}$, contains not only $\partial_{a}$ but also a term like 
 $\vartheta^{a}\partial_{\mu}$. From $\vartheta^{a}\partial_{\mu}=(\vartheta^{a}\star)\,\partial_{\mu} -\frac{1}{2}\Omega^{a A}\partial_{A}\partial_{\mu}$, it is clear that the supersymmetry operator does not satisfy the Leibniz rule anymore, and hence the supersymmetry is generically broken on the noncommutative superspace. Nevertheless, as we will show in the following section, for the `minimally noncommutative chiral superspace',    the full supersymmetry can remain  unbroken.\\

The integration over the chiral superspace reads
 \begin{equation}
 \displaystyle{\int dx^{D}d\theta^{N/2}W_{\star}(x,\theta)}
 =\displaystyle{\sqrt{\frac{(2\pi)^{D}\det{\theta}}{\det{\tilde{c}}}}\,
 \Tr\left[\Gamma_{\!\!\theta}\O(W_{\star})\right]}\,,
 \end{equation}
where $\Gamma_{\!\!\theta}$ is a normalized matrix,\footnote{Alternatively one can take  the anti-symmetric product of $\hpartial_{\alpha}$'s as $\Gamma_{\!\!\theta}$ with a suitable normalization constant, which would pick up the coefficient of the $\theta^{N/2}$ term only even for the integration over the generic nonchiral superfield, $W(x,\theta,\bar{\theta})$. However, restricting on the  chiral superpotential,  $W(x,\theta)$, of which the  component field for the highest order in $\theta$ is bosonic,  the above choice of $\Gamma_{\!\!\theta}$ serves the desired property well.}
\begin{equation}
 \ba{ll}
\Gamma_{\!\!\theta}=
\displaystyle{{\frac{\hbtheta{}^{[N/2}\cdots\hbtheta{}^{2}\hbtheta{}^{1]}}{\sqrt{\det\tilde{c}\,}}}}\,,
~~~&~~~\tr\left(\Gamma_{\!\theta}(\Gamma_{\!\theta})^{\dagger}\right)=1\,.\ea
 \end{equation}

A remarkable fact about  Eq.(\ref{action}) is that   there exists a fine tuned commutative limit, since
$\mbox{sdet}{\,\Omega}=\det(i\theta)/\det{s}$ can remain fixed. This also implies that  the Jacobian factor
in the partition function expression (\ref{partition}) converges to a nonzero constant, which would not be true
for  the bosonic noncommutative theory.     In the commutative limit, the potential remains fixed but the kinetic term diverges as it contains the derivatives. The prescription for the supermatrix partition function is then to take the saddle point analysis with the constraint on the supermatrices such that the divergent part of the kinetic term vanishes.  Similarly for the chiral theories, one can take the commutative limit where   $\det(i\theta)/\det{\tilde{c}}$ remains fixed and impose   a  constraint on  the supermatrices which prevents the kinetic term from diverging.  In this way, there exits a  supermatrix description  for any superfield theory on a commutative superspace too. The  supermatrix model  Lagrangian is  same as that of the superfield theory, but the supermatrices  are surely of infinite size meaning the planar limit. The dual particles for the superfields in even and odd  dimensions are D-instantons and D0-branes respectively to be consistent  with the T-duality.

\section{Examples\label{examples}}
\subsection{$\N=1$ $D=3$ nonlinear sigma model}
In the Minkowskian three dimensions, the $\N=1$ superspace consists of  $\{x^{\mu},\vartheta^{\alpha}\}$ where $\alpha=1,2$ and the odd coordinates are Majorana spinors, $(\vartheta^{\alpha})^{\dagger}=\vartheta^{\alpha}$. The generic nonlinear sigma model  reads with the real superfields, $\Phi^{i}=\phi^{i}(x)+i\vartheta^{\alpha}\psi^{i}_{\alpha}(x)+i\vartheta^{1}\vartheta^{2}F^{i}(x)$, \cite{Gates:nr}
\begin{equation}
\displaystyle{\int dx^{3}d\vartheta^{2}\,-\half g_{ij}(\Phi)D^{\alpha}\Phi^{i}D_{\alpha}\Phi^{j}}\,,
\end{equation}
where $D_{\alpha}=\partial_{\alpha}+(\epsilon\gamma^{\mu}\vartheta)_{\alpha}\partial_{\mu}$, $D^{\alpha}=\epsilon^{-1 \alpha\beta}D_{\beta}$ and $\epsilon$ is the $2\times 2$ anti-symmetric matrix satisfying $\epsilon\gamma^{\mu}\epsilon^{-1}=-(\gamma^{\mu})^{T}$. The supersymmetry transformation is given by
$\delta\Phi^{i}=\varepsilon^{\alpha} Q_{\alpha}\Phi^{i}$ where $\varepsilon^{\alpha}$ is a real spinor and $Q_{\alpha}=\partial_{\alpha}-(\epsilon\gamma^{\mu}\vartheta)_{\alpha}\partial_{\mu}$.

From (\ref{orditostar}) we  note  $D_{\alpha}\Phi^{i}=\partial_{\alpha}\Phi^{i}+
\half\{(\epsilon\gamma^{\mu}\vartheta)_{\alpha},\partial_{\mu}\Phi^{i}\}_{\star}$. With this expression we replace all the ordinary products by the star products to turn on the noncommutativity on the superspace,
\begin{equation}
\ba{cc}
[\x^{1},\x^{2}]=i\lambda^{2}\,,~~~&~~~
\{\hvartheta^{\alpha},\hvartheta^{\beta}\}=2\lambda^{2}\delta^{\alpha\beta}\,,
\ea
\end{equation}
where $\lambda$ is a real parameter we introduce to control the strength of the noncommutativity so that $\x^{1}\sim\x^{2}\sim\hvartheta^{\alpha}\sim\lambda$.   Note that the time coordinate, $t$, commutes with all other coordinates.   In particular, using the pauli matrices we set $\hvartheta^{1}=\lambda\sigma^{1}$, $\hvartheta^{2}=\lambda\sigma^{2}$. The action on the noncommutative superspace reads
\begin{equation}
\displaystyle{\int \!dx^{3}d\vartheta^{2}\,-\half g_{\star ij}(\Phi)\star(D^{\alpha}\Phi^{i})\star(D_{\alpha}\Phi^{j})=i\pi\half\!\int\!dt\,\mbox{Str}\!\left[
g_{ij}(\hat{\Phi})(\hat{D}{}^{\alpha}\hat{\Phi}{}^{i})(\hat{D}_{\alpha}\hat{\Phi}{}^{j})\right]}\,,
\end{equation}
where $\mbox{Str}$ denotes the supertrace, $\Tr[\sigma^{3}\,\cdot\,]$, and  from (\ref{zwmap}) $\Phi^{j}$'s are generically time dependent Hermitian supermatrices. Their covariant derivatives are of the form, 
\begin{equation}
\hat{D}_{\alpha}\hat{\Phi}=\half\lambda^{-1}[\sigma^{\alpha},\hat{\Phi}]+
\half\lambda(\epsilon\gamma^{0})_{\alpha\beta}\{\sigma^{\beta},\partial_{0}\hat{\Phi}\}+
\half(\epsilon\gamma^{l})_{\alpha\beta}\{\sigma^{\beta},[\hat{n}_{l},\hat{\Phi}]\}\,,
\label{DPhi}
\end{equation}
where $\hat{n}_{1}=i\x^{2}/\lambda$,  $\hat{n}_{2}=-i\x^{1}/\lambda$ are the normalized harmonic oscillators, $[\hat{n}_{1},\hat{n}_{2}]=-i$. 

The supersymmetry is surely completely broken. From (\ref{orditostar}), (\ref{zwmap}) and (\ref{consistency}) with the rescaling,   $\varepsilon\rightarrow 2\lambda\varepsilon\sigma^{3}$, the broken supersymmetry is given by
\begin{equation}
\delta_{broken}\hat{\Phi}=[\varepsilon^{\alpha}\sigma^{3}\sigma^{\alpha},\hat{\Phi}]-
\lambda^{2}(\epsilon\gamma^{0})_{\alpha\beta}\{\varepsilon^{\alpha}\sigma^{3}\sigma^{\beta},
\partial_{0}\hat{\Phi}\}-
\lambda(\epsilon\gamma^{i})_{\alpha\beta}\{\varepsilon^{\alpha}\sigma^{3}\sigma^{\beta},
[\hat{n}_{i},\hat{\Phi}]\}\,.
\label{Q3d}
\end{equation}

In the commutative limit, $\lambda\rightarrow 0$, from (\ref{DPhi})  the kinetic term tends to diverge. The saddle point prescription requires then $[\sigma^{\alpha},\hat{\Phi}^{i}]=0$, and hence $\hat{\Phi}^{i}=\hat{\phi}^{i}(\x)\otimes 1_{2\times 2}$. At the saddle point the action reduces to the bosonic matrix action,
\begin{equation}
\displaystyle{-2\pi\!\int\!dt\,\mbox{tr}\!\left(
g_{ij}(\hat{\phi})[\hat{n}_{l},\hat{\phi}{}^{i}][\hat{n}_{l},\hat{\phi}{}^{j}]\right)}\,.
\end{equation}
From (\ref{Q3d}), the supersymmetry is trivially ``restored" as $\delta\hat{\Phi}=i[\varepsilon^{\alpha}\sigma_{\alpha},\hat{\Phi}]=0$.

\subsection{$\N=1$ $D=4$ chiral theories - unbroken SUSY}
In the Minkowskian four dimensions, the  $\N$-extended superspace reads $\{x^{\mu},\theta^{a\alpha},\bar{\theta}_{a}^{\dalpha}\}$, where $\bar{\theta}_{a}^{\dalpha}=(\theta^{a\alpha})^{\dagger}$, $\alpha,\dalpha=1,2$ and $1\leq a\leq\N$ so that the dimension of the  fermionic real coordinates is $N=4\N$. The covariant derivatives are 
\begin{equation}
\ba{ll}
D_{a\alpha}=\partial_{a\alpha}+i(\sigma^{\mu}\bar{\theta}_{a})_{\alpha}\partial_{\mu}\,,~~~~&~~~~
\bar{D}^{a}_{\dalpha}=\bar{\partial}_{\dalpha}^{a}+i(\theta^{a}\sigma^{\mu})_{\dalpha}\partial_{\mu}\,.\ea
\end{equation}
The supersymmetry transformation is given by the following real differential operator acting on any superfield,
\begin{equation}
i(\varepsilon^{a}Q_{a}+\bar{Q}^{a}\bar{\varepsilon}_{a})=i
\varepsilon^{a\alpha}\partial_{a\alpha}-i\bar{\varepsilon}^{\dalpha}_{a}\bar{\partial}_{\dalpha}^{a}
+(\theta^{a}\sigma^{\mu}\bar{\varepsilon}_{a}+\varepsilon^{a}\sigma^{\mu}\bar{\theta}_{a})\partial_{\mu}\,,
\end{equation}
where $\varepsilon^{a\alpha}$, $\bar{\varepsilon}_{a}^{\dalpha}=(\varepsilon^{a\alpha})^{\dagger}$ are the supersymmetry transformation parameters.  \\

The conventional chiral  superfield is defined to satisfy  $\bar{D}_{\dalpha}^{a}\Phi=0$  so that it is a function of  $x_{+}^{\mu}=x^{\mu}+i\theta^{a}\sigma^{\mu}\bar{\theta}_{a}$ and $\theta^{a\alpha}$ only, $\Phi(x_{+},\theta)$. However, in order to incorporate the noncommutativity into the chiral superspace  such that   the star product preserves the `chirality',  it is necessary to shift  the complex variables, $x^{\mu}_{+}$ to the real ones,  $x^{\mu}$, and take  $\Phi(x,\theta)$ as the newly defined ``chiral" superfield satisfying $\bar{\partial}_{\dalpha}^{a}\Phi=0$. We define the shifting operator,
\begin{equation}
J=\displaystyle{e^{-i\theta^{a}\sigma^{\mu}\bar{\theta}_{a}\partial_{\mu}}}\,,
\end{equation}
to satisfy $J^{\dagger}=J^{-1}$, $J\Phi(x_{+},\theta)=\Phi(x,\theta)$, $J(\Phi\Phi^{\prime}) =(J\Phi)(J\Phi^{\prime})$,  and 
$J\bar{D}_{\dalpha}^{a}J^{\dagger}=\bar{\partial}_{\dalpha}^{a}$. 
The supersymmetry transformation of the chiral superfield is then,
\begin{equation}
\ba{ll}
\delta\Phi(x,\theta)=\Q\Phi(x,\theta)\,,~~&~\Q\equiv
iJ(\varepsilon^{a}Q_{a}+\bar{Q}^{a}\bar{\varepsilon}_{a})J^{\dagger}= 
i\varepsilon^{a\alpha}\partial_{a\alpha}-i\bar{\varepsilon}^{\dalpha}_{a}\bar{\partial}_{\dalpha}^{a}
+2\theta^{a}\sigma^{\mu}\bar{\varepsilon}_{a}\partial_{\mu}\,.\ea\label{Qexp}
\end{equation}
Similarly the anti-chiral superfield, $\bar{\Phi}(x,\bar{\theta})=J^{\dagger}\bar{\Phi}(x_{-},\bar{\theta})$, transforms as
\begin{equation}
\ba{ll}
\delta\bar{\Phi}(x,\bar{\theta})=\bar{\Q}\bar{\Phi}(x,\bar{\theta})\,,~~&~\bar{\Q}\equiv
iJ^{\dagger}(\varepsilon^{a}Q_{a}+\bar{Q}^{a}\bar{\varepsilon}_{a})J= 
i\varepsilon^{a\alpha}\partial_{a\alpha}-i\bar{\varepsilon}^{\dalpha}_{a}\bar{\partial}_{\dalpha}^{a}
+2\varepsilon^{a}\sigma^{\mu}\bar{\theta}_{a}\partial_{\mu}\,.\ea \label{Qbarexp}
\end{equation}

As an example we consider the $\N=1$ chiral scalar  theory \cite{Wess:cp}. Acting $J$ on the standard Lagrangian  gives rise to only a total derivative term. Thus, we can rewrite the chiral superfield action in a different but equivalent form which is ready to incorporate the noncommutativity on the superspace.\footnote{This kind of rewriting the action  was first considered in   \cite{Kawai:2003yf} which also includes the gauge multiplet. Our point here is to show that the full supersymmetry can be unbroken for the  scalar theory on the minimally noncommutative superspace. Including the gauge multiplet appears to break the supersymmetry. It would be very interesting to find a mechanism not to break the supersymmetry in that case too.}  From $J[\Phi(x_{+},\theta)\bar{\Phi}(x_{-},\bar{\theta})]=\Phi(x,\theta)J^{2}\bar{\Phi}(x,\bar{\theta})$, the action reads
\begin{equation}
\displaystyle{\int dx^{4}d\theta^{2}d\bar{\theta}^{2}\,\Phi(x,\theta)J^{2}\bar{\Phi}(x,\bar{\theta})+
\int dx^{4}d\theta^{2}\,W(\Phi(x,\theta))+\int dx^{4}d\btheta^{2}\,\overline{W}(\bar{\Phi}(x,\btheta))}\,,
\end{equation}
where  explicitly,
\begin{equation}
J^{2}=1-2i\theta\sigma^{\mu}\btheta\partial_{\mu}-\theta^{2}\bar{\theta}^{2}\Box\,.\label{Jsquare}
\end{equation} 
From $\Q J^{2}=J^{2}\bar{\Q}$, the supersymmetry is manifest. \\

Before replacing the ordinary products by the star products, it is crucial to note that from (\ref{orditostar}) we   can rewrite $J^{2}\bar{\Phi}(x,\bar{\theta})$ in terms of the star product,
\begin{equation}
J^{2}\bar{\Phi}=\bar{\Phi}-
\half i\sigma^{\mu}_{\alpha\dalpha}[\theta^{\alpha},\{\btheta^{\dalpha},\partial_{\mu}\bar{\Phi}\}_{\star}]_{\star}-
\textstyle{\frac{1}{16}}[\theta^{\alpha},\{\theta_{\alpha},[\btheta_{\dalpha},\{\btheta^{\dalpha},\Box\bar{\Phi}
\}_{\star}]_{\star}\}_{\star}]_{\star}\,.
\end{equation}
With this understanding, we insert the star products into the action such that $\Phi\star(J^{2}\bar{\Phi})$, $W_{\star}(\Phi)$, $\overline{W}_{\star}(\bar{\Phi})$. Up to the total derivative term, the kinetic term remains   same as in the commutative case, while the potential terms  generically get nontrivial star product deformations but maintaining   the `chiral' property.  The map to the supermatrix model is then straightforward. We merely mention that the size of the gamma matrices for the odd coordinates must be $4\times 4$ for the nondegenerate case, $\det S\neq 0$ resulting in the generic supermatrix model ({\textit{cf.}}~\cite{Kawai:2003yf}). The notion of  `chiral' supermatrix is  valid as 
$[\lambda^{2}\hat{\bar{\partial}}_{\dalpha},\hat{\Phi}]=0$.\\

From (\ref{Qexp}) and (\ref{Qbarexp}) it is clear that the supersymmetry is  broken completely  for the generic case,  $c^{\alpha\beta}=(\bar{c}^{\dalpha\dbeta})^{\ast}\neq 0$. However, for the minimally noncommutative superspace, where $c^{\alpha\beta}=0$,  as $\Q$ satisfies the Leibniz rule acting on  the `chiral'  superfields, the full supersymmetry is unbroken.\footnote{For the invariance of the bosonic theory under the scaling of the noncommutativity parameter, see \cite{Ishikawa:2001mq}.}\\

To consider the commutative limit, we take  the  noncommutative chiral superspace, (\ref{chiral}), with the replacement, $(\theta^{\mu\nu},c^{\alpha\beta},\tilde{c}^{\alpha\dbeta},\bar{c}^{\dalpha\dbeta})\rightarrow
(\lambda^{2}\theta^{\mu\nu},  \lambda^{4}c^{\alpha\beta},    \lambda^{4}\tilde{c}^{\alpha\dbeta},
\lambda^{4}\bar{c}^{\dalpha\dbeta})$. In the commutative limit, $\lambda\rightarrow 0$,
$\x^{\mu}\sim\lambda$, $\htheta^{\alpha}\sim\lambda^{2}$,  $\partial_{\mu}\sim\lambda^{-1}$,
$\partial_{\alpha}\sim\lambda^{-2}$, but
$\det(\lambda^{2}\theta)/\det(\lambda^{4}\tilde{c})=\det(\theta)/\det(\tilde{c})$ remains fixed.
Thus, the kinetic part diverges while the superpotential term  is finite. Calculating the  partition function  (\ref{partition}), it is  natural to use the saddle point approximation with  the constraint on the supermatrix, 
$\partial^{\alpha}\partial_{\alpha}\hat{\Phi}=\partial_{\alpha}\partial_{\mu}\hat{\Phi}=0$, and hence $\hat{\Phi}=1_{4\times 4}\hat{\phi}(\x/\lambda)+\lambda^{-2}\hat{\theta}^{\alpha}\Gamma_{\!\!(5)}\psi_{\alpha}$ where $\psi_{\alpha}$ is a constant Grassmann variable. The whole kinetic term vanishes then in the supermatrix model.  The supersymmetry is restored, from (\ref{deriv}), (\ref{orditostar}),  (\ref{zwmap}), (\ref{consistency}), (\ref{Qexp}), as   the adjoint action, 
\begin{equation}
\delta\hat{\Phi}=i\left[{{\small{ \left(\ba{cc}\varepsilon^{\alpha}&0\,\\0~&-\varepsilon^{\alpha}\ea\right)}}}
\lambda^{2}\hat{\partial}_{\alpha}\,,\,\hat{\Phi}\right]\,,
\end{equation}
which preserves the constraint, $\partial^{\alpha}\partial_{\alpha}\hat{\Phi}=\partial_{\alpha}\partial_{\mu}\hat{\Phi}=0$.

\section{Comments\label{comments}}
It is straightforward to generalize the  above analysis  to the $\N$-extended supersymmetry in the Minkowskian four dimensions.  The commutative limit keeping the chiral superpotential finite is given by    $\x^{\mu}\sim\lambda^{\N}$, $\htheta^{\alpha}\sim\lambda^{2}$ so that
$Q_{\alpha}\sim\lambda^{-2}(1+\lambda^{4-\N})$,
where the zeroth order inside the bracket comes from $\partial_{\alpha}$ while the $({4-\N})$th order is
from the term, $\theta\partial_{\mu}$,  which breaks the Leibniz rule.  The overall factor, $\lambda^{-2}$,
can be absorbed into the supersymmetry parameter.  Thus only for $\N=1,2,3$ cases the supersymmetry can be  restored for the dual supermatrix model in the limit. This is consistent with the rigidity of the  $\N=4$ theory or the absence of the  superpotential in the theory.  Furthermore, the commutative limit to keep the integral over the full superspace finite is  $\x^{\mu}\sim\lambda^{\N}$, $\htheta^{\alpha}\sim\lambda$ so that
$Q_{\alpha}\sim\lambda^{-1}(1+\lambda^{2-\N})$. Thus, the $\N=4$ case would not restore the supersymmetry again.  
We take this as an evidence  that there should  be no $\N=4$ superfield formalism. 
Similarly one can argue that there should be no non-chiral $\N=2$ superfield formalism in four dimensions.\\


\acknowledgments{The author wishes to thank  J-H Cho, K. Lee, S-K Nam, S-J Rey and  S-H Yi for the valuable
discussions.  This work is the result of research activities,  Astrophysical Research Center for the Structure and
Evolution of the Cosmos,   supported by Korea Science $\&$ Engineering Foundation.}

\newpage
\appendix
\section{Appendix~:~the proof of the isomorphism}
Here we prove the isomorphism (\ref{iso}) between the operator and the star product formalisms  by the mathematical induction on $n$  with $f_{n}=z^{A_{1}}\cdots z^{A_{n}}$. Firstly we note for arbitrary $m\geq 0$,   $g_{m}=z^{B_{1}}\cdots z^{B_{m}}$,
\begin{equation}
\ba{l}
\z^{A}\O(g_{m})-\O(z^{A}g_{m})\\
{}\\
=\displaystyle{\sum_{P=1}^{m!}\frac{(-1)^{N_{P}}}{(m+1)!}\sum_{j=1}^{m+1}
\left[\z^{A}\z^{P_{1}}\z^{P_{2}}\cdots\z^{P_{m}}-(-1)^{\#_{\!A}(\#_{\!P_{1}}+\cdots +\#_{\!P_{j-\!1}})}
\z^{P_{1}}\cdots\z^{P_{j-\!1}}\z^{A}\z^{P_{j}}\cdots\z^{P_{m}}\right]}\\
{}\\
=\displaystyle{\sum_{P=1}^{m!}\frac{(-1)^{N_{P}}}{(m+1)!}\sum_{j=1}^{m}\,
[\z^{A},\z^{P_{1}}\cdots\z^{P_{j}}\}\z^{P_{j+\!1}}\cdots\z^{P_{m}}}\\
{}\\
=\displaystyle{\sum_{P=1}^{m!}\frac{(-1)^{N_{P}}}{(m+1)!}\sum_{j=1}^{m}
\left(\sum_{l=1}^{j}(-1)^{\#_{\!A}\!(\#_{\!P_{1}}+\cdots+\#_{\!P_{l-\!1}})}\z^{P_{1}}\cdots \z^{P_{l-\!1}}[\z^{A},\z^{P_{l}}\}\z^{P_{l+\!1}}\cdots \z^{P_{j}}\right)\z^{P_{j+\!1}}\cdots\z^{P_{m}}}\\
{}\\
=\displaystyle{\sum_{P=1}^{m!}\frac{(-1)^{N_{P}}}{(m+1)!}\sum_{j=1}^{m}(m+1-j)
(-1)^{\#_{\!A}\!(\#_{\!P_{1}}+\cdots+\#_{\!P_{j-\!1}})}\z^{P_{1}}\cdots \z^{P_{j-\!1}}[\z^{A},\z^{P_{j}}\}\z^{P_{j+\!1}}\cdots \z^{P_{m}}}\\
{}\\
=\displaystyle{\sum_{P=1}^{m!}\frac{(-1)^{N_{P}}}{(m+1)!}\sum_{j=1}^{m}(m+1-j)
(-1)^{\#_{\!P_{j}}\!(\#_{\!P_{1}}+\cdots+\#_{\!P_{j-\!1}})}[\z^{A},\z^{P_{j}}\}
\z^{P_{1}}\cdots {\not\!\z^{P_{j}}}\cdots \z^{P_{m}}}\\
{}\\
=\textstyle{\frac{1}{2}}\displaystyle{\sum_{k=1}^{m}(-1)^{\#_{\!B_{k}}\!(\#_{\!B_{1}}+\cdots+\#_{\!B_{k-\!1}})}
[\z^{A},\z^{B_{k}}\}\O(z^{B_{1}}\cdots{\not\!z^{B_{k}}}\cdots z^{B_{m}})}\\
{}\\
=\displaystyle{\O(\half\Omega^{AC}\partial_{C}g_{m})}\,.
\end{array}
\end{equation}
This shows $\O(\z^{A})\O(g)=\O(z^{A}\star g)$ for arbitrary $g$. Now we assume $\O(f_{k})\O(g)=\O(f_{k}\star g)$ for $k\leq n-1$. 
From 
\begin{equation}
\ba{ll}
\O(f_{n})=\frac{1}{n}{\displaystyle{\sum_{j=1}^{n}}}\,\varepsilon_{\!j}\,
\z^{A_{j}}\O(z^{A_{1}}\cdots{\not\!z^{A_{j}}}\cdots z^{A_{n}})\,,~~&~~ \varepsilon_{j}=(-1)^{\#_{\!A_{j}}(\#_{\!A_{1}}+\cdots+\#_{\!A_{j-\!1}})}\,,
\ea
\end{equation}
it follows
\begin{equation}
\O(f_{n})\O(g)=\frac{1}{n}\displaystyle{{\sum_{j=1}^{n}}\,\varepsilon_{\!j}\,\O\left(
z^{A_{j}}\star[(z^{A_{1}}\cdots {\not\!z^{A_{j}}}\cdots z^{A_{n}})\star g]\right)}\,.\label{tocombine}
\end{equation}
On the other hand, a straightforward manipulation gives for any $1\leq j\leq n$,
\begin{equation}
\ba{l}
(z^{A_{1}}\cdots z^{A_{n}})\left(\half\lrOmega\right)^{m}g\\
{}\\
=\varepsilon_{\!j}(z^{A_{j}}z^{A_{1}}\cdots{\not\!z^{A_{j}}}\cdots z^{A_{n}})\left(\half\lrOmega\right)^{m}g\\
{}\\
=\varepsilon_{\!j}\left[\ba{l}
z^{A_{j}}\left\{(z^{A_{1}}\cdots{\not\!z^{A_{j}}}\cdots z^{A_{n}})\left(\half\lrOmega\right)^{m}g\right\}\\
{}\\
+m\half\Omega^{A_{j}B}\partial_{B}\left\{
(z^{A_{1}}\cdots{\not\!z^{A_{j}}}\cdots z^{A_{n}})\left(\half\lrOmega\right)^{m-1}g\right\}\\
{}\\
-m\half\displaystyle{\sum_{l< j}}\,\varepsilon_{\!l}\Omega^{A_{j}A_{l}}\left\{
(z^{A_{1}}\cdots{\not\!z^{A_{l}}}\cdots{\not\!z^{A_{j}}}\cdots z^{A_{n}})\left(\half\lrOmega\right)^{m-1}g\right\}\\
{}\\
-m\half\displaystyle{\sum_{l>j}}\,\varepsilon_{\!l}(-1)^{\#_{\!A_{l}}\#_{\!A_{j}}}\Omega^{A_{j}A_{l}}\left\{
(z^{A_{1}}\cdots{\not\!z^{A_{j}}}\cdots{\not\!z^{A_{l}}}\cdots z^{A_{n}})\left(\half\lrOmega\right)^{m-1}g\right\}\ea\right]\,.
\ea
\end{equation}
From (\ref{OmegaOmega}), summing over $j$ cancels the last two terms so that 
\begin{equation}
(z^{A_{1}}\cdots z^{A_{n}})\star g=
\frac{1}{n}\displaystyle{\sum_{j=1}^{n}}~\varepsilon_{\!j}(z^{A_{j}}+\half\Omega^{A_{j}B}\partial_{B})\left\{
(z^{A_{1}}\cdots{\not\!z^{A_{j}}}\cdots z^{A_{n}})\star g\right\}\,.
\end{equation}
Thus, combining with (\ref{tocombine}) we get $\O(f_{n})\O(g)=\O(f_{n}\star g)$,  which completes our proof.


\newpage
\begin{thebibliography}{99}
\bibitem{BFSS}
T. Banks, W. Fischler, S.H. Shenker and L. Susskind, Phys. Rev. D {\bf 55} (1997) 5112.


\bibitem{Ishibashi:1996xs}
N.~Ishibashi, H.~Kawai, Y.~Kitazawa and A.~Tsuchiya, 
Nucl.\ Phys.\ B {\bf 498} (1997) 467.


\bibitem{Taylor:2001vb}
W.~Taylor,
Rev.\ Mod.\ Phys.\  {\bf 73} (2001) 419
[hep-th/0101126].



\bibitem{Douglas:2001ba}
M.~R.~Douglas and N.~A.~Nekrasov,
Rev.\ Mod.\ Phys.\  {\bf 73} (2001) 977
[hep-th/0106048].

\bibitem{JHPpp}
S.~Hyun and J.-H.~Park,
JHEP {\bf 0211} (2002) 001
[hep-th/0209219];\\
S.~Hyun, J.-H.~Park and S.~H.~Yi,
JHEP {\bf 0303} (2003) 004
[hep-th/0301090].




\bibitem{Poly}
A.P. Polychronakos,  
``Generalized statistics in one dimension,''
[hep-th/9902157];\\
A.P. Polychronakos, Phys.\ Lett.\ B {\bf 266}, 29 (1991).





\bibitem{JHP}
J.-H.~Park, Phys. Lett. A  {\bf 307} (2003) 183.

\bibitem{largeN1}
T. Eguchi and H. Kawai, Phys. Rev. Lett. {\bf 48}  (1982) 1063;\\
G. Parisi, Phys. Lett. B {\bf 112} (1982) 463;\\
D. Gross and Y. Kitazawa, Nucl. Phys. B {\bf 206} (1982) 440;\\
G. Bhanot, U. Heller and H. Neuberger, Phys. Lett. B {\bf 113} (1982) 47;\\
S. Das and S. Wadia, Phys. Lett. B {\bf 117} (1982) 228.

\bibitem{largeN2}
A. Gonzales-Arroyo and M. Okawa, Phys. Rev. D {\bf 27} (1983) 2397.



\bibitem{DV}
R.~Dijkgraaf and C.~Vafa,
Nucl.\ Phys.\ B {\bf 644} (2002) 3;\\
R.~Dijkgraaf and C.~Vafa,
Nucl.\ Phys.\ B {\bf 644} (2002) 21;\\
R.~Dijkgraaf and C.~Vafa, ``A perturbative window into non-perturbative physics,'' [hep-th/0208048].




\bibitem{Klemm:2001yu}
D.~Klemm, S.~Penati and L.~Tamassia,
Class.\ Quant.\ Grav.\  {\bf 20} (2003) 2905.


\bibitem{Ooguri:2003qp}
H.~Ooguri and C.~Vafa, ``The C-deformation of gluino and non-planar diagrams,'' [hep-th/0302109].

\bibitem{Seiberg:2003yz}
N.~Seiberg,
JHEP {\bf 0306} (2003) 010.






\bibitem{deBoer:2003dn}
J.~de Boer, P.~A.~Grassi and P.~van Nieuwenhuizen, ``Non-commutative superspace from string theory,'' [hep-th/0302078].

\bibitem{Kawai:2003yf}
H.~Kawai, T.~Kuroki and T.~Morita,
``Dijkgraaf-Vafa theory as large-N reduction,''
[hep-th/0303210].

\bibitem{Chepelev:2003ga}
I.~Chepelev and C.~Ciocarlie, ``A path integral approach to noncommutative superspace,'' [hep-th/0304118].

\bibitem{Britto:2003aj}
R.~Britto, B.~Feng and S.~J.~Rey, ``Deformed superspace, N = 1/2 supersymmetry and (non)renormalization  theorems,''
[hep-th/0306215].

\bibitem{Terashima:2003ri}
S.~Terashima and J.~T.~Yee, ``Comments on Noncommutative Superspace,'' [hep-th/0306237].




\bibitem{Hatsuda:2003ry}
M.~Hatsuda, S.~Iso and H.~Umetsu, ``Noncommutative Superspace, Supermatrix and Lowest Landau Level,'' [hep-th/0306251].

\bibitem{Ferrara}
S.~Ferrara, M.~A.~Lledo and O.~Macia, ``Supersymmetry in noncommutative superspaces,'' [hep-th/0307039].










\bibitem{Buchbinder:uq}
I.~L.~Buchbinder and S.~M.~Kuzenko, ``Ideas and Methods of Supersymmetry and Supergravity: A Walk through Superspace,''
IOP Publishing (1995).



\bibitem{commentJHP}
D.~Bak, K.~Lee and J.-H. Park, Phys. Lett. B {\bf 501} (2001) 305.

\bibitem{Langmann:2003cg}
E.~Langmann, R.~J.~Szabo and K.~Zarembo, 
``Exact solution of noncommutative field theory in background magnetic  fields,'' [hep-th/0303082].

\bibitem{Gates:nr}
S.~J.~Gates, M.~T.~Grisaru, M.~Rocek and W.~Siegel,
Front.\ Phys.\  {\bf 58} (1983) 1.


\bibitem{Wess:cp}
J.~Wess and J.~Bagger, ``Supersymmetry And Supergravity,'' Princeton Univ. Pr. (1992).


\bibitem{Ishikawa:2001mq}
T.~Ishikawa, S.~I.~Kuroki and A.~Sako,
J.\ Math.\ Phys.\  {\bf 43} (2002) 872. 



\end{thebibliography}
\end{document}